\documentclass[notitlepage,letterpaper, 11pt]{article}
\usepackage[ansinew]{inputenc} 
\usepackage[colorlinks=true,urlcolor=blue,linkcolor=blue,citecolor=blue]{hyperref} 
\usepackage{graphicx}
\usepackage{amsfonts}
\usepackage{amssymb}
\usepackage{color}
\usepackage{harvard}
\usepackage{cite}

\usepackage{amsthm}

\begin{document}

\title {Are there any models with \\ homogeneous energy density?}
\author{
J. Ospino\thanks{j.ospino@usal.es}, J.L. Hern\'andez-Pastora \thanks{jlhp@usal.es} \\
\textit{Departamento de Matem\'atica Aplicada and} \\
\textit{Instituto Universitario de F\'isica Fundamental y Matem\'aticas,}  \\ \textit{Universidad de Salamanca, Salamanca-Spain;} \\
H. Hern\'andez\thanks{hector@ula.ve} and L.A. N\'u\~nez\thanks{lnunez@uis.edu.co} \\
\textit{Escuela de F\'isica, Universidad Industrial de Santander, } \\ \textit{ Bucaramanga-Colombia and} \\ \textit{Departamento de F\'isica, Universidad de Los Andes,} \\ \textit{ M\'erida-Venezuela.}
}
\maketitle

\begin{abstract}
By applying a recent method --based on a tetrad formalism in General Relativity and the orthogonal splitting of the Riemann tensor--  to the simple spherical static case, we found that the only static solution with homogeneous energy density is the Schwarzschild solution and  that there are no spherically symmetric dynamic  solutions  consistent with the homogeneous energy density assumption. Finally, a circular equivalence is shown among the most frequent conditions considered in the spherical symmetric case: homogeneous density, isotropy in pressures, conformally flatness and shear-free conditions. We  demonstrate that, due to the regularity conditions at the center of the matter distribution, the imposition of two conditions necessarily leads to the static case.
\end{abstract} 
\maketitle

\section{Introduction}
The case of a uniform density spherical matter configuration is all but the standard entry point in all textbooks of General Relativity and Relativistic Astrophysics \cite{Weinberg1972,ShapiroTeukolsky1983,Schutz2009,MisnerThorneWheeler2017}, presenting the most ``simple'' interior Schwarzschild solution. Despite its physical inconsistency --it models a fluid with an infinite sound speed-- its simplicity is of a pedagogical value in illustrating the methods used in solving physical systems in different (static \& dynamic) interesting scenarios \cite{Wyman1949,BonnorFaulkes1967,MisraSrivastava1973,BowersLiang1974,Ponce1986,HerreraNunez1987,MaharajMaartens1989,HerreraNunez1989,NunezRueda2007}.    

In a recent paper \cite{OspinoHernandezNunez2017}, by using a tetrad formalism in General Relativity and the orthogonal splitting of the Riemann tensor, we proposed a full set of equations  equivalent to the Einstein system which governs the evolution of self-gravitating systems. The formalism was applied to the spherical case to show, through a very simple static case, that it is possible to obtain relevant information from these self-gravitating systems. 

The study of the geometric and kinematic properties of timelike congruences is fundamental in the analysis of the evolution of self-gravitating fluids and there it is common the use of a framework based on the known 1+3 formalism \cite{Ellis1971, EllisBruni1989, Ehlers1993, WainwrightEllis2005,EllisVanElst1999,TsagasChallinorMaartens2008,EllisMaartensMacCallum2012}. In this formalism, any tensor quantity can be split into components along a tangent vector to a timelike congruence, and in its corresponding components orthogonal to it.

The method we used consists of constructing two sets of independent equations, which contain the same information as the Einstein equations, expressed in terms of scalar functions. 

As a starting point we choose an orthogonal unitary tetrad, and the first set of evolution equations is obtained from the projection of the Riemann tensor along the unit tetrad. This is equivalent to the use of Ricci identities, which will allow us to define the physical variables and the scalars of the Weyl tensor.
The second set of six constrain equations is obtained directly from  the Bianchi identities. For the spherical case, solving this system of first order equations will provide us with the necessary information to know the equation of state of a gravitational source with spherical symmetry.

Using this method we shall obtain three results involving isotropic and anisotropic solutions to the Einstein Equations with homogeneous energy density. First we show that the only static solution with homogeneous energy density is the Schwarzschild isotropic solution. Secondly, it is shown there are no spherically symmetric dynamic solution consistent with homogeneous energy density and, for this case the shear-free assumption is equivalent to the isotropic pressure condition. Finally, a circular equivalence is shown among the most frequent conditions considered for the spherical symmetric case: homogeneous density, isotropy in pressures, conformally flatness \cite{BWStewart1982,JWainwright1984,GoodeWrinwright1985,WBBonor1985,WBBonor1987,JPonce1988,GoodeColeyWaiwright,HDOF2001,LHerrera2003,GeneralStudy,GronSteinar,ManjonjoMaharajMoopanar2017,ManjonjoMaharajMoopanar2018,BVIvanov} and shear-free conditions \cite{CBCollinsJWainwright1983,ENGlass1979,HStephaniTWolf1996,GovinderGovenderMaarteens1998,JoshiDadhichMaartens,PankaRituparnoGoswamiNaresh,LHerreraNSantos2003, LHerreraNSantosAWang,LHerreraADiPriscoJOspino2014}. It is  shown that due to the regularity conditions at the center of the matter distribution, the imposition of any two of them necessarily leads to the static case.

The paper is organized as follows: In section 2 we present the general strategy and the formalism in obtaining all relevant equations. Section 3 develops the analysis for the spherical case and also considers the particular spherical-static case. Finally, in Section 4 we present the final conclusions.  

\section{The strategy and general formalism}
As we have mentioned above, the strategy we shall follow is to compile two independent sets of equations, expressed in terms of scalar functions, which contain the same information as the Einstein system.

Let us choose an orthogonal unitary tetrad:
\begin{equation}
\label{TetradGen}
e^{(0)}_\alpha~=~V_\alpha, \, e^{(1)}_\alpha~=~K_\alpha, \, e^{(2)}_\alpha~=~L_\alpha \; \; \mathrm{and}  \; e^{(3)}_\alpha~=~S_\alpha.
\end{equation}

As usual, $\eta_{(a)(b)}~=~ g_{\alpha\beta} e_{(a)}^\alpha e_{(b)}^\beta$, with $a=0,\,1,\,2,\,3$, i.e. latin indices label different vectors of the tetrad. Thus, the tetrad satisfies the standard relations:
\begin{eqnarray}
V_{\alpha}V^{\alpha} &=& -K_{\alpha}K^{\alpha} = -L_{\alpha}L^{\alpha} = -S_{\alpha}S^{\alpha} = -1\,, \nonumber \\
V_{\alpha}K^{\alpha} &=& V_{\alpha}L^{\alpha} = V_{\alpha}S^{\alpha} = K_{\alpha}L^{\alpha} = K_{\alpha}S^{\alpha} = S_{\alpha}L^{\alpha} = 0\,.  \nonumber
\end{eqnarray} 
With the above tetrad (\ref{TetradGen}) we shall also define the corresponding directional derivatives operators
\begin{equation}
\label{DirectionalDerivatives}
f^{\bullet} = V^{\alpha} \partial_{\alpha}f ; \quad f^{\dag} = K^{\alpha} \partial_{\alpha}f  \quad \mathrm{and} \quad f^{\ast} = L^{\alpha}\partial_{\alpha}f .
\end{equation}

The first set can be considered  purely geometrical and emerges from the projection of the Riemann tensor along the tetrad \cite{Wald2010}, i.e.
\begin{eqnarray}
\label{RiemannProj}
2V_{\alpha\, ;[\beta ;\gamma]} = R_{\delta \alpha \beta \gamma} V^{\delta} \, , &\quad &  \nonumber 2K_{\alpha\, ;[\beta ;\gamma]} = R_{\delta \alpha \beta \gamma} K^{\delta} \,, \nonumber  \\
 & &  \\
2L_{\alpha \, ;[\beta ;\gamma]} = R_{\delta \alpha \beta \gamma} L^{\delta} \,  &\quad \mathrm{and} \quad &  \nonumber 2S_{\alpha\, ;[\beta ;\gamma]} = R_{\delta \alpha \beta \gamma} S^{\delta}\,  ; \nonumber
\end{eqnarray}
where $e^{(a)}_{\alpha \, ;\beta \gamma}$ are the second covariant derivatives of each tetrad (\ref{tetrad}) vector indicated with $a = 0,1,2,3.$

The second set emerges from the Bianchi identities:
\begin{equation}
\label{BianchiIdent}
R_{\alpha \beta [\gamma \delta \, ;  \mu]} =
R_{\alpha \beta \gamma \delta \, ;  \mu}  + R_{\alpha \beta  \mu \gamma  \, ; \delta} + R_{\alpha \beta \delta \mu \, ; \gamma } = 0 \,.
\end{equation}

\section{Spherical Case}
In this section we shall present the relevant equations, for the spherically symmetric, locally anisotropic, dissipative, collapsing matter configuration, written in terms of the kinematical variables: the four acceleration $a_\alpha$, the expansion scalar $\Theta$, the shear tensor $\sigma$ and some scalars functions (the structure scalars related to the splitting of the Riemann Tensor).

\subsection{The tetrad, the source and kinematical variables}

To proceed with the above objective we shall restrict to a spherically symmetric line element given by
\begin{equation}
\mathrm{d}s^2=-A^2\mathrm{d}t^2+B^2 \mathrm{d}r^2+R^2 (\mathrm{d}\theta ^2+\sin^2(\theta)\mathrm{d}\phi ^2)\, ,
\label{SphericMetric}
\end{equation}
where the coordinates are: $x^0=t$, $x^1=r$, $x^2=\theta$, and $x^3=\phi$; with $A(t,r)$, $B(t,r)$ and $R(t,r)$ functions of their arguments.

For this case the tetrad can written as:
\begin{eqnarray}
\label{tetrad}
V_{\alpha} &=& \left(-A,0,0,0\right),\,
K_{\alpha}=\left(0,B,0,0\right), \,
L_{\alpha}=\left(0,0,R,0\right), \nonumber \\
&\mathrm{and}& \quad S_{\alpha} =
\left(0,0,0,R \sin(\theta)\right) \,. 
\end{eqnarray}
The covariant derivatives of the orthonormal tetrad are:
\begin{eqnarray}
V_{\alpha;\beta}&=&-a_1K_\alpha V_\beta+\sigma_1K_\alpha K_\beta+\sigma_2(L_\alpha L_\beta+S_\alpha S_\beta),  \nonumber \\
K_{\alpha;\beta}&=&-a_1V_\alpha V_\beta+\sigma_1 V_\alpha K_\beta+J_1(L_\alpha L_\beta+S_\alpha S_\beta), \nonumber \\
L_{\alpha;\beta}&=&\sigma_2 V_\alpha L_\beta-J_1K_\alpha L_\beta+J_2S_\alpha S_\beta
\quad \mathrm{and} \label{CovDTetrad} \\
S_{\alpha;\beta}&=&\sigma_2 V_\alpha S_\beta-J_1K_\alpha S_\beta -J_2L_\alpha S_\beta \,. \nonumber
\end{eqnarray}
Where: $J_1$, $J_2$, $\sigma_{1}$,  $\sigma_{2}$ and $a_1$ are expressed in terms of the metric functions and their derivatives as:
\begin{eqnarray}
\label{metricquantities}
J_1=\frac{1}{B}\frac{R^{\prime}}{R} \,,\,\, 
J_2&=&\frac{1}{R}\cot(\theta)\,,\,\,
\sigma_{1} = \frac{1}{A}\frac{\dot{B}}{B}\,,\,\,
\sigma_{2} = \frac{1}{A}\frac{\dot{R}}{R} \nonumber \\
&\mathrm{and}& \quad  a_1=\frac{1}{B}\frac{A^{\prime}}{A} \,, 
\end{eqnarray}
with primes and dots representing radial and time derivatives, respectively.

As we mentioned before we shall assume  our source as a bounded, spherically symmetric, locally anisotropic, dissipative, collapsing matter configuration, described by a general energy momentum tensor, written in the ``canonical'' form, as:
\begin{equation}
{T}_{\alpha\beta}= (\rho+P) V_\alpha V_\beta+P g _{\alpha \beta} +\Pi_{\alpha \beta}+\mathcal{F}_\alpha V_\beta+\mathcal{F}_\beta V_\alpha .
\label{EnergyTensor}
\end{equation}

It is immediately seen that the physical variables can be defined --in the Eckart frame  where fluid elements are at rest-- as:
\begin{eqnarray}
\rho &=& T_{\alpha \beta} V^\alpha V^\beta, \, \mathcal{F}_\alpha=-\rho V_\alpha - T_{\alpha\beta}V^\beta,\,\, 
P = \frac{1}{3} h^{\alpha \beta} T_{\alpha \beta} \nonumber \\
&\mathrm{and}& \quad \Pi_{\alpha \beta} = h_\alpha^\mu h_\beta^\nu \left(T_{\mu\nu} - P h_{\mu\nu}\right)\,,
\end{eqnarray}
with $h_{\mu \nu}=g_{\mu\nu}+V_\nu V_\mu$.

As can be seen from the condition $\mathcal{F}^{\mu} V_{\mu}=0$, and the symmetry of the problem, Einstein Equations imply $T_{03}=0$, thus:
\begin{equation}
\mathcal{F}_{\mu}=\mathcal{F}K_{\mu} \quad \Leftrightarrow \quad \mathcal{F}_{\mu}= \left(0, \frac{\mathcal{F}}{B}, 0, 0\right) .
\label{energyflux}
\end{equation}
Clearly $\rho$ is the energy density (the eigenvalue of $T_{\alpha\beta}$ for eigenvector $V^\alpha$), $\mathcal{F}_\alpha$ represents the  energy flux four vector;  $P$ corresponds to the isotropic pressure, and $\Pi_{\alpha \beta}$ is the anisotropic tensor, which can be  expressed as
\begin{equation}
\Pi_{\alpha \beta}= \Pi_{1}\left(K_\alpha K_\beta -\frac{h_{\alpha \beta}}{3}\right)
\label{anisotropictensor},
\end{equation}
with
\begin{equation}
\Pi_{1}=\left(2K^{\alpha} K^{\beta} +L^{\alpha} L^{\beta}\right)  T_{\alpha \beta}.
\label{Pi1}
\end{equation}
Finally, we shall express the kinematical variables (the four acceleration, the expansion scalar and the shear tensor) for a self-gravitating fluid as:
\begin{eqnarray}
a_\alpha&=&V^\beta V_{\alpha;\beta}=a K_\alpha =\left(0, \frac {A^{\prime} }{A },0,0\right),
\label{aceleration} \\
\Theta&=&V^\alpha_{;\alpha} =\frac{1}{A}\,\left(\frac{\dot{B}}{B}+\frac{2\dot{R}}{R}\right), \label{theta} \\
\sigma&=&\frac{1}{A}\left(\frac{\dot{B}}{B} -\frac{\dot{R}}{R}\right)\,.  \label{shear}
\end{eqnarray}

\subsection{The orthogonal splitting of the Riemann tensor and structure scalars}

In this section we shall introduce a set of scalar functions --the structure scalars--  obtained from the orthogonal splitting of the Riemann tensor (see \cite{GarciaParrado2007,HerreraEtal2009B,HerreraEtal2014}) which has proven to be very useful in expressing the Einstein Equations.

Following \cite{GarciaParrado2007}, we can express the splitting of the Riemann tensor as:
\begin{eqnarray}
R_{\alpha \beta \mu \nu}&=&2V_\mu V_{[\alpha}Y_{\beta] \, \nu}+2h_{\alpha[\nu}X_{\mu] \,  \beta}+2V_\nu V_{[\beta}Y_{\alpha] \, \mu}\nonumber
\\
&+& h_{\beta\nu}(X_0 \, h_{\alpha\mu}-X_{\alpha\mu})+h_{\beta\mu}(X_{\alpha\nu} -X_0 \, h_{\alpha\nu}) \nonumber \\
&+& 2V_{[\nu} Z_{ \, \mu]}^{\gamma}\varepsilon_{{\alpha \beta \gamma}} +2V_{[\beta} Z_{{\,  \ \alpha]}}^{{\gamma }}\ \varepsilon_{{\mu \nu \gamma}} \,,
\end{eqnarray}
with $\varepsilon_{\mu \nu \gamma} = \eta_{\phi \mu \nu \gamma} V^{\phi}$, and  $ \eta_{\phi \mu \nu \gamma}$ the Levi-Civita 4-tensor. The corresponding Ricci contraction for the above Riemann tensor can also be written as:
\begin{eqnarray}
R_{\alpha\mu}&=& Y_0 \, V_\alpha V_\mu-X_{\alpha \mu}-Y_{\alpha\mu} +X_0 \, h_{\alpha\mu}  +Z^{\nu \beta} \varepsilon_{\mu \nu \beta}V_{\alpha} \nonumber \\
&&+V_{\mu} Z^{\nu \beta} \varepsilon_{\alpha \nu \beta} \, ;
\end{eqnarray}
where the quantities: $Y_{\alpha\beta}$, $X_{\alpha\beta}$ and $Z_{\alpha\beta}$ can be expressed as
\begin{eqnarray}
Y_{\alpha\beta}&=&\frac{1}{3}Y_0 \, h_{\alpha\beta} +Y_1\left[K_\alpha K_\beta-\frac{1}{3} h_{\alpha\beta}\right] \,, \\
X_{\alpha\beta}&=&\frac{1}{3} X_0 \, h_{\alpha\beta} +X_1\left[K_\alpha K_\beta-\frac{1}{3} h_{\alpha\beta}\right] \\
\mathrm{and} & & \nonumber \\
Z_{\alpha\beta}&=&Z \, (L_\alpha S_\beta-L_\beta S_\alpha)\,,
\end{eqnarray}
with
\begin{eqnarray}
Y_0 &=& 4\pi(\rho+3P)\,, \,\,  
Y_1=\mathcal{E}_1-4\pi \Pi_1 \,, \,\,
X_0=8\pi \rho\,,\nonumber \\
X_1 &=& -(\mathcal{E}_1+4\pi \Pi_1) \quad
\mathrm{and} \quad  Z= 4 \pi \mathcal{F} \,,\label{varfis}
\end{eqnarray}
and the electric part of the Weyl tensor is written as
\begin{equation}
E_{\alpha\beta}=C_{\alpha\nu\beta\delta}V^\nu V^\delta = \mathcal{E}_1\left[K_\alpha K_\beta-\frac{1}{3} h_{\alpha\beta}\right]\,.
\end{equation}

\subsection{Projections of Riemann Tensor}

From the above system (\ref{RiemannProj}) (by using the covariant derivative of equations (\ref{CovDTetrad}) and the projections of the orthogonal splitting of the Riemann tensor) we can obtain the  first set of independent equations, for the spherical case, in terms of  $J_1$, $J_2$, $\sigma_{1}$,  $\sigma_{2}$, and $a_1$, (defined in (\ref{metricquantities})) and their directional derivatives, i.e.
\begin{eqnarray}
\sigma^{\bullet}_{1} -a_1^\dag-a_1^2+\sigma_1^2&=&-\frac{1}{3}(Y_0+2Y_1) \, , \label{ecR1} \\
\sigma^{\bullet}_{2} +\sigma_2^2-a_1J_1&=&\frac{1}{3}(Y_1-Y_0) \, , \label{ecR2}  \\
\sigma_2^\dag+J_1(\sigma_2-\sigma_1)&=&-Z \, , \label{ecR3} \\
J^{\bullet}_{1} +J_1\sigma_2-a_1\sigma_2&=&-Z \, , \label{ecR4} \\
J_1^\dag+J_1^2-\sigma_1 \sigma_2&=&\frac{1}{3}(X_1-X_0) \, , \label{ecR5} \\
J^{\bullet}_{2} +J_2\sigma_2&=&0 \, , \label{ecR6} \\
J_2^\dag+J_1J_2&=&0 \qquad \mathrm{and}  \label{ecR7} \\
J_1^2-\frac{1}{R^2}-\sigma_2^2&=&-\frac{1}{3}(X_0+2X_1) \,. \label{ecR8}
\end{eqnarray}

\subsection{Equations from Bianchi identities}
The second set of equations for the spherical case, emerge from the independent Bianchi identities (\ref{BianchiIdent}), and can be written as:
\begin{eqnarray}
a_1[-X_0+X_1&-&Y_0+Y_1]+3J_1Y_1+3Z^{\bullet} \nonumber \\
\qquad \qquad +6Z \sigma_1&+&3Z\sigma_2-Y_0^\dag+Y_1^\dag=0 \,,
\label{Bianchi1} \\
X^{\bullet}_0 -X^{\bullet}_1 -6a_1Z &-&3J_1Z  +\left[Y_0 -Y_1 -X_1\right] \sigma_1  
\nonumber \\
 +\left[Y_0 +2Y_1 -X_1\right] \sigma_2 &+&X_0[\sigma_1+\sigma_2]-3Z^\dag=0 \,, \label{Bianchi2} \\
X^{\bullet}_0 +2 X^{\bullet}_1 &+&2X_0\sigma_2 -6J_1Z \nonumber \\
+[4X_1 &+& 2Y_0 - 2Y_1] \sigma_2=0, \label{Bianchi3} \\  
X_0^\dag+2X_1^\dag&+&6J_1X_1+6Z\sigma_2=0 \,. \label{Bianchi4}
\end{eqnarray}

\subsection{The static case}

In the line element  (\ref{SphericMetric}) we can assume, without any loss of generality, $R=r$ and integrate (\ref{ecR2}) to obtain:
\begin{equation}
A=C_1 e^{\int B^2 r(Y_0-Y_1)dr} \,.
\label{Aest1}
\end{equation}
Next, from (\ref{ecR4}) it follows at once that:
\begin{equation}
B^2=\frac{1}{1-\frac{r^2}{3}(X_0+2X_1)}\label{BestB1}\,,
\end{equation}
where $C_1$ is a constant of integration. These metric elements (\ref{Aest1}) and (\ref{BestB1}) expressed in terms of the structure scalars $X_1$  and $Y_0-Y_1$, describe any static anisotropic sphere (see reference \cite{HerreraOspinoDiPrisco2008}).

\subsection{Models with homogeneous energy density}
It is easy to check that the Schwarzschild interior solution corresponds to the case $X_1=Y_1=0$, and it follows clearly from (\ref{varfis}) that $\mathcal{E}_1 = \Pi _1 = 0$.

Now, let us show that if we have the homogeneous energy density premise, the only possible outcome is the isotropic Schwarzschild solution. Thus, let us consider  models with homogeneous energy density
\begin{equation}
X_0=8\pi \rho=\mbox{cte.} \,,
\label{denconst}
\end{equation}
and study the consequences derived from this assumption, under certain physically reasonable circumstances. First, taking into account (\ref{denconst}) and integrating equation (\ref{Bianchi4}), we obtain
\begin{equation}
X_1=\frac{C}{r^3}\label{X1C}.
\end{equation}
Next consider the regularity condition at $r= 0$ via (\ref{X1C}), which implies $C=0$, then:
\begin{equation}
\mathcal{E}_1+4\pi \Pi_1=0.\label{EPi}
\end{equation}

Also from equation (\ref{Bianchi1}), taking into account (\ref{ecR2}) and (\ref{BestB1}) we get
\begin{equation}
P^\prime _r=-4\pi r\frac{(P^2_r+\frac{4}{3}\rho P_r+\frac{1}{3}\rho ^2)}{1-\frac{8\pi}{3}\rho r^2}-\frac{2}{r}\Pi_1 \, . \label{TOV}
\end{equation}
Clearly, if  the anisotropic term $(\Pi _1)$ is zero at a point other than the origin, it will be zero at all points \cite{HerreraSantos1997}. If $\Pi _1(r)$ does not vanish it must be positive or negative and, from equation (\ref{EPi}),  the same thing will be true for Weyl's scalar $\mathcal{E}_1$. But, given that $\mathcal{E}_1(0)=0$, by the conditions of regularity at the origin, and $\mathcal{E}_1(R_{\Sigma})=P_t >0$, by the boundary conditions, we find that $\mathcal{E}_1(r)>0$.  On the other hand, if $\Pi_1=-\Delta <0$, from (\ref{TOV}) we find that there is a  $ r_c< r_\Sigma$ given by
\begin{equation}
r^2_c=\frac{3 \Delta }{2\pi (3P^2_r+4\rho P_r+\rho ^2+4\rho \Delta)},\label{rcri}
\end{equation}
for which $P^\prime _r=0$, showing that the minimum radial pressure is reached in a $r_c$ smaller than $r_\Sigma$. Thus, we conclude that $\Pi_1(r)$ and $\mathcal{E}_1(r)$ must vanish, i.e.
\begin{equation}
\Pi_1(r) >0 \; \mathrm{and}  \; \mathcal{E}_1(r)>0\,\,\Rightarrow \, \Pi_1(r)=\mathcal{E}_1(r) =0.
\end{equation}

Therefore, we can see that the only static solution with homogeneous energy density, under the above considerations, is necessarily the Schwarzschild solution. More over, if we  require that the circumference $2\pi R$ of an infinitesimal sphere about the origin be just $2\pi$ times its proper radius $Bdr$,
that is \cite{MisnerSharp1964}
\begin{equation}
B(t,r)=R^\prime (t,r)\,\, \mathrm{when}\,\, r\rightarrow\,0 \,. 
\label{ECond}
\end{equation}
\noindent Now, replacing (\ref{ECond}) into (\ref{ecR8}), in the static case, we get
\begin{equation}
X_0(0)=0!!!\label{EX0}
\end{equation}
From (\ref{EX0}) we conclude that the models with homogeneous energy density do not satisfy the euclidean condition (\ref{ECond}).

\subsection{The  non-static case}
\subsubsection{Regularity on the origin}
To guarantee a good asymptotic behavior of the metric (\ref{SphericMetric}), in the vicinity of the origin, we must demand that the functions $A(r,t)$, $B(r,t)$ and $R(r,t)$, have the following analytical form, from the standard Taylor expansion:
\begin{eqnarray}
\label{conReg}
A(r,t)&=&\alpha _0+\alpha _1(t) r+\alpha _2 (t) r^2+ \cdots \nonumber\\
B(r,t)&=&1+\beta _1(t) r+\beta_2 (t)r^2+ \cdots \\
R(r,t)&=&r+\gamma _1(t)r^2+ \gamma _2(t)r^3+ \cdots \nonumber 
\end{eqnarray}

\subsubsection{ The Case $X_{0}=X_{0} (t)$ and $Z=0$ }
In this case, we obtain from equation (\ref{Bianchi4}) that $X_1=0$ and equations (\ref{Bianchi1})-(\ref{Bianchi3}) become:
\begin{eqnarray}
a_1[Y_1&-&X_0-Y_0]+3J_1Y_1=[Y_0-Y_1]^\dag \,, \label{Bianchi1R} \\
X_0^{\bullet}+[X_0&+&Y_0-Y_1]\sigma_1+[X_0+Y_0+2Y_1]\sigma_2=0\,, \label{Bianchi2R}\\
X_0^{\bullet}&+&2[X_0+Y_0-Y_1]\sigma_2=0\label{Bianchi3R}\,.
\end{eqnarray}
Notice that if $X_0=$cte.,then from the equation (\ref{Bianchi3R}), it follows that $\sigma _2=0$, and we get the static case (iii) analyzed in \cite{OspinoHernandezNunez2017}.

Next, combining equations (\ref{Bianchi2R}) and (\ref{Bianchi3R}) we find that
\begin{equation}
[Y_0+X_0-Y_1][\sigma_1-\sigma_2]+3Y_1\sigma_2=0,
\label{eqdensidad}
\end{equation}
or
\begin{equation}
[\rho +P_r][\sigma_1-\sigma_2]=2\Pi_1\sigma_2,
\label{eqdensidad1}
\end{equation}
where from equation (\ref{varfis}) and $X_1=0$.
 From the equation (\ref{eqdensidad1}), it follows that
\begin{equation}
\sigma_1-\sigma_2=0\,\,\Leftrightarrow\,\, \Pi_1=0\label{rhocon}.
\end{equation} 

In other words, the shear-free and isotropic pressure  conditions are equivalent, for  non-dissipative fluids with homogeneous energy density.

Evaluating equation (\ref{Bianchi3R}) at $r=0$ and taking into account (\ref{conReg}) we obtain
\begin{equation}
  \sigma _1=\sigma _2=0\,\Rightarrow \,X_0=\mbox{cte.} \,,
  \label{dcte} 
\end{equation}
therefore it follows that there are no spherically symmetric dynamic solutions with homogeneous energy density.

\subsection{Circular conditions }
In this section we shall prove the equivalence of the following circular conditions taken two by two:
\begin{itemize}
\item Homogeneous energy density, $X_0=X_0(t)$ \textcircled{1} 
\item Isotropy in the pressures, $\Pi_1=0$ \textcircled{2} 
\item Conformally flat, $\mathcal{E}=0$ \textcircled{3}
\item Shear-free condition, $\sigma_1=\sigma_2$ \textcircled{4} 
\end{itemize}

\paragraph{\textcircled{1} and \textcircled{2} $\leftrightarrow$ \textcircled{3} and  \textcircled{4}.} If we assume \textcircled{1} and \textcircled{2}, then from (\ref{Bianchi4}) we find that
$X_1=0 \,\Rightarrow \, Y_1=0 \Rightarrow $\textcircled{3}, and by using equation (\ref{eqdensidad}) we get
$\sigma_1=\sigma_2, \Rightarrow$ \textcircled{4}

On the other hand, if we assume \textcircled{3} and \textcircled{4}, with the result obtained in \cite{OspinoHernandezNunez2017},
\begin{equation}
\sigma_1=\sigma_2 \Rightarrow a_1=0,
\label{previousResult}
\end{equation}
and from the subtraction of (\ref{ecR1}) from (\ref{ecR2}), we find
$Y_1=X_1=0\,\Rightarrow$~\textcircled{2}, and from (\ref{eqdensidad}) we obtain
$X_0=X_0(t) \Rightarrow$~\textcircled{1}.

\paragraph{\textcircled{1} and \textcircled{3} $\leftrightarrow$ \textcircled{2}  and \textcircled{4}.}  Now if we assume \textcircled{1} and \textcircled{3},  we find from (\ref{Bianchi4}) that
$X_1=0 \Rightarrow  Y_1=0 \Rightarrow$~\textcircled{2}. Again, by using equation (\ref{eqdensidad}) we get $\sigma_1=\sigma_2, \Rightarrow$~\textcircled{4}. On the other hand, if we assume \textcircled{2}  and \textcircled{4}, using (\ref{previousResult}) and again, substracting (\ref{ecR1}) from (\ref{ecR2}), we find $Y_1=X_1=0\,\Rightarrow$ \textcircled{3}, now with (\ref{eqdensidad}), we obtain $X_0=X_0(t)\,\Rightarrow$ \textcircled{1}.

\paragraph{\textcircled{1} and \textcircled{4} $\leftrightarrow$ \textcircled{2}  and  \textcircled{3}.} If we assume\textcircled{1} and \textcircled{4}, then from (\ref{Bianchi4}) and (\ref{eqdensidad}) we obtain
$X_1=Y_1=0 \Rightarrow$ \textcircled{2} and \textcircled{3}. On the other hand, if we assume \textcircled{2} and \textcircled{3},then
\begin{equation}
X_1=Y_1=0, \label{X1Y1}
\end{equation}
by substituting (\ref{X1Y1}) into (\ref{Bianchi4}) we obtain
\begin{equation}
X_0=X_0(t)\Rightarrow \textcircled{1} \,. \label{X0}
\end{equation}
 Next, from (\ref{X0}) and by replacing  (\ref{X1Y1}) in (\ref{eqdensidad}) we find 
$\sigma_1=\sigma_2\Rightarrow$ \textcircled{4}. 
Notice that if we also take into account (\ref{conReg}), in all cases previously considered, we only get the static case.

\section{Final Remarks}
We have found that, despite its simplicity and pedagogical interest, the uniform density spherical matter configuration is a very restricted and unphysical solution to the Einstein Equations. 

As we have stated above in this short paper we have presented several results concerning the homogeneous energy density assumption for isotropic and anisotropic solutions to the  Einstein Equations. First, we have shown that if the regularity condition at the center of the distribution and some other physical reasonable boundary condition at the surface of the distribution are to be satisfied, then the only static solution for a spherically symmetric matter distribution with homogeneous energy density is the Schwarzschild isotropic solution. This rules out any anisotropic generalization for $\rho = \mbox{const}$ found in the literature  \cite{BowersLiang1974, MaharajMaartens1989} and complements the proof for the classic problem that a static perfect fluid star should be spherically symmetric for physically reasonable isotropic equation of state \cite{Lindblom1988,LindblomMasood1994, Masood2007,Pfister2011}. More over, we have shown that even for the static homogeneous  Schwarzschild solution the center of the matter distribution has to be excluded because it does not satisfy the Euclidean condition. Clearly, is possible to obtain viable solutions if this condition is relaxed assuming a core-envelope model (see \cite{TakisaMaharaj2016} and references therein). 

Secondly, it is shown there are no spherically symmetric dynamic solution consistent  with homogeneous energy density and, for this case the shear-free assumption is equivalent to the isotropic pressure condition. 

Finally, we have considered the most frequent conditions assumed in a spherical symmetric case: homogeneous density, isotropy in pressures, conformally flatness and shear-free conditions (see  \cite{Wyman1949,BonnorFaulkes1967,MisraSrivastava1973,BowersLiang1974,Ponce1986,HerreraNunez1987,MaharajMaartens1989,HerreraNunez1989,NunezRueda2007,HerreraOspinoDiPrisco2008,lake2003all,rahman2002spacetime,HerreraDiPriscoOspino2010,Herrera:2001vg} and reference therein). It is found that the two of these assumptions are necessarily equivalent to the other remainder  two.  Additionally, it is  demonstrated that, due to the regularity conditions at the center of the matter distribution,  the imposition of two of them necessarily leads to the static case.

Again, we have shown that the most simple and ``pedagogic'' spherical matter solution --$\rho = \mbox{const}$--  is very restricted and unphysical, but there has  been much recent work with variable energy densities,  satisfying all physical criteria, that seems to correspond to more realistic matter configurations \cite{MatondoMaharajRay2018,MauryaMaharaj2018}.

\section*{Acknowledgments}
J.O. and J.L.H.-P.  acknowledge financial support from Fondo Europeo de Desarrollo
Regional (FEDER) (grant FIS2015-65140-P) (MINECO/FEDER). J.O acknowledges hospitality of School of Physics of the Industrial University of Santander, Bucaramanga Colombia. L.A.N. gratefully acknowledge the financial support of the Vicerrector\'ia de Investigaci\'on y Extensi\'on de la Universidad Industrial de  Santander and the financial support provided by COLCIENCIAS under Grant No. 8863   


\end{document}